\documentclass[aps,prb,twocolumn]{revtex4}
\usepackage{makeidx}
\usepackage{amsmath,amssymb,amsfonts,amsthm}
\usepackage{graphicx,bm}

\begin{document}

\title{Piezoelectric softening in ferroelectrics: ferroelectric versus
antiferroelectric PbZr$_{1-x}$Ti$_{x}$O$_{3}$}
\date{}
\author{F. Cordero,$^{1}$ F. Craciun,$^{1}$ F. Trequattrini$^{2}$ and C.
Galassi$^{3}$}
\affiliation{$^1$ CNR-ISC, Istituto dei Sistemi Complessi, Area della Ricerca di Roma -
Tor Vergata,\\
Via del Fosso del Cavaliere 100, I-00133 Roma, Italy}
\affiliation{$^{2}$ Dipartimento di Fisica, Universit\`{a} di Roma \textquotedblleft La
Sapienza\textquotedblright , P.le A. Moro 2, I-00185 Roma, Italy}
\affiliation{$^{3}$ CNR-ISTEC, Istituto di Scienza e Tecnologia dei Materiali Ceramici,
Via Granarolo 64, I-48018 Faenza, Italy}

\begin{abstract}
The traditional derivation of the elastic anomalies associated with
ferroelectric (FE) phase transitions in the framework of the Landau theory
is combined with the piezoelectric constitutive relations instead of being
explicitly carried out with a definite expression of the FE part of the free
energy. In this manner it is shown that the softening within the FE phase is
of electrostrictive and hence piezoelectric origin. Such a piezoelectric
softening may be canceled by the better known piezoelectric stiffening, when
the piezoelectric charges formed during the vibration are accompanied by the
depolarization field, as for example in Brillouin scattering experiments. As
experimental validation, we present new measurements on Zr-rich PZT, where
the FE phase transforms into antiferroelectric on cooling or doping with La,
and a comparison of existing measurements made on FE PZT with low frequency
and Brillouin scattering experiments.
\end{abstract}

\maketitle


\section{Introduction}

The structural ferroelastic transitions are accompanied by elastic softening,%
\cite{Sal90} and the same is true for most ferroelectric (FE),\cite{SL98}
magnetic and electronic transitions whose order parameter is coupled to
strain.\cite{Lut07} The softening starts in the high temperature phase when
approaching the transition temperature, and can be directly or indirectly
associated with a soft mode, or precursor fluctuations. In the low
temperature phase, the elastic constants generally partially or completely
restiffen, partly due to the freezing of the fluctuations, or even become
much stiffer than in the high temperature phase, as in systems where a
charge disproportionation and/or charge ordering transitions freezes the
dynamic Jahn-Teller fluctuations.\cite{CTB11} The FE and antiferroelectric
(AFE) transitions are also ferroelastic and therefore are accompanied by
sizeable softening, but the shape of the compliance curves versus
temperature strongly depends on the material and also on the measurement
frequency and technique. An additional extrinsic softening mechanism in the
FE state, especially important at the lowest frequencies, is the motion of
the domain walls.\cite{ZS96,SSK09}

It is argued that the main intrinsic contribution to the softening within
the FE phase is of electrostrictive and hence piezoelectric origin, but can
be completely canceled by the so-called piezoelectric stiffening when the
electric depolarization fields accompanying the piezoelectric charge can
fully develop. As examples where these phenomena are particularly evident,
we present new dielectric and elastic measurements on Zr-rich PZT, where the
FE phase can be turned into AFE by cooling or doping with La, and compare
existing low frequency\cite{CTC13} and Brillouin scattering\cite{KKK12}
experiments on PZT at the morphotropic phase boundary between the
rhombohedral and tetragonal FE phases.

\section{Experimental and Results}

Ceramic samples of Pb$_{1-3x/2}$La$_{x}$Zr$_{1-y}$Ti$_{y}$O$_{3}$ (PLZT $%
100x/100\left( 1-y\right) /100y$) with $y=$ 0.046 and $x=$ 0, 0.02 were
prepared by the mixed-oxide method in the same manner as a previous series
of samples.\cite{127,145} The oxide powders were calcined at 700~${^{\circ }}
$C for 4 hours, pressed into bars, sintered at 1250~${{}^{\circ }}$C for 2~h
and packed with PbZrO$_{3}$\ + 5wt\% excess ZrO$_{2}$\ to prevent PbO loss.
The powder X-ray diffraction analysis did not reveal any trace of impurity
phases. The densities were about 95\% of the theoretical values and the
grains were large, with sizes of $5-20$~$\mu $m. For the anelastic
experiments samples were cut as a thin bars $4~$cm long and $0.6$~mm thick,
and electrodes were applied with Ag paint. The Young's modulus or its
reciprocal, the compliance $s$, was measured by suspending the bar on two
thin thermocouple wires and electrostatically exciting the flexural
resonance.\cite{135} The compliances are shown normalized to their values $%
s_{0}$ in the paraelectric (PE) phase, in terms of the resonance frequency $%
f\sim 1$~kHz: $s\left( T\right) /s_{0}=$ $\left[ f_{0}/f\left( T\right) %
\right] ^{2}$. The dielectric permittivity $\epsilon $ was measured on discs
with diameters of 12~mm and 0.7~mm thick by means of a HP~4284A LCR meter
with a four-wire probe and an electric field of 0.5 V/mm, between 0.2 and
200~kHz.

According to the phase diagram proposed by Asada and Koyama\cite{AK04} for
PLZT $x/95/5$, below $T_{\mathrm{C}}$ the La-free sample becomes
rhombohedral (R) FE, while below $T_{\mathrm{AF}}$ it becomes orthorhombic
(O) antiferroelectric (AFE), but for $x\left( \text{La}\right) >$ 0.01 the
intermediate phase is incommensurate (IC) AFE instead of R-FE, with reduced $%
T_{\mathrm{C}}$ and enhanced $T_{\mathrm{AF}}$. In our samples the Zr/Ti
ratio is very close to 95/5 and both the dielectric and elastic measurements
conform to the phase diagram of Ref. \onlinecite{AK04}. Figure \ref%
{fig_AnDiPiezo} presents the permittivity $\epsilon $ measured at 1~kHz and
the normalized compliance $s/s_{0}$ measured at $\sim 1.7$~kHz during
heating. When looked from high temperature, both permittivity curves show
the Curie-Weiss rise up to $T_{\mathrm{C}}$, followed by a drop due to the
first order nature of the FE and IC-AFE transition, and an additional
decrease below $T_{\mathrm{AF}}$, where the susceptibility in a fully
ordered AFE state is quite reduced with respect to a FE or IC state. While
there is no qualitative difference between the $\epsilon $ curves of the
sample with intermediate R-FE and IC-AFE phases, the difference is
outstanding in the $s$ curves: the FE phase has an additional roughly
constant softening, indicated as a hatched area, completely missing in the
IC-AFE phase, so that the step at $T_{\mathrm{AF}}$ is even inverted in sign.

\section{Discussion}

\begin{figure}[tbh]
\includegraphics[width=8.5 cm]{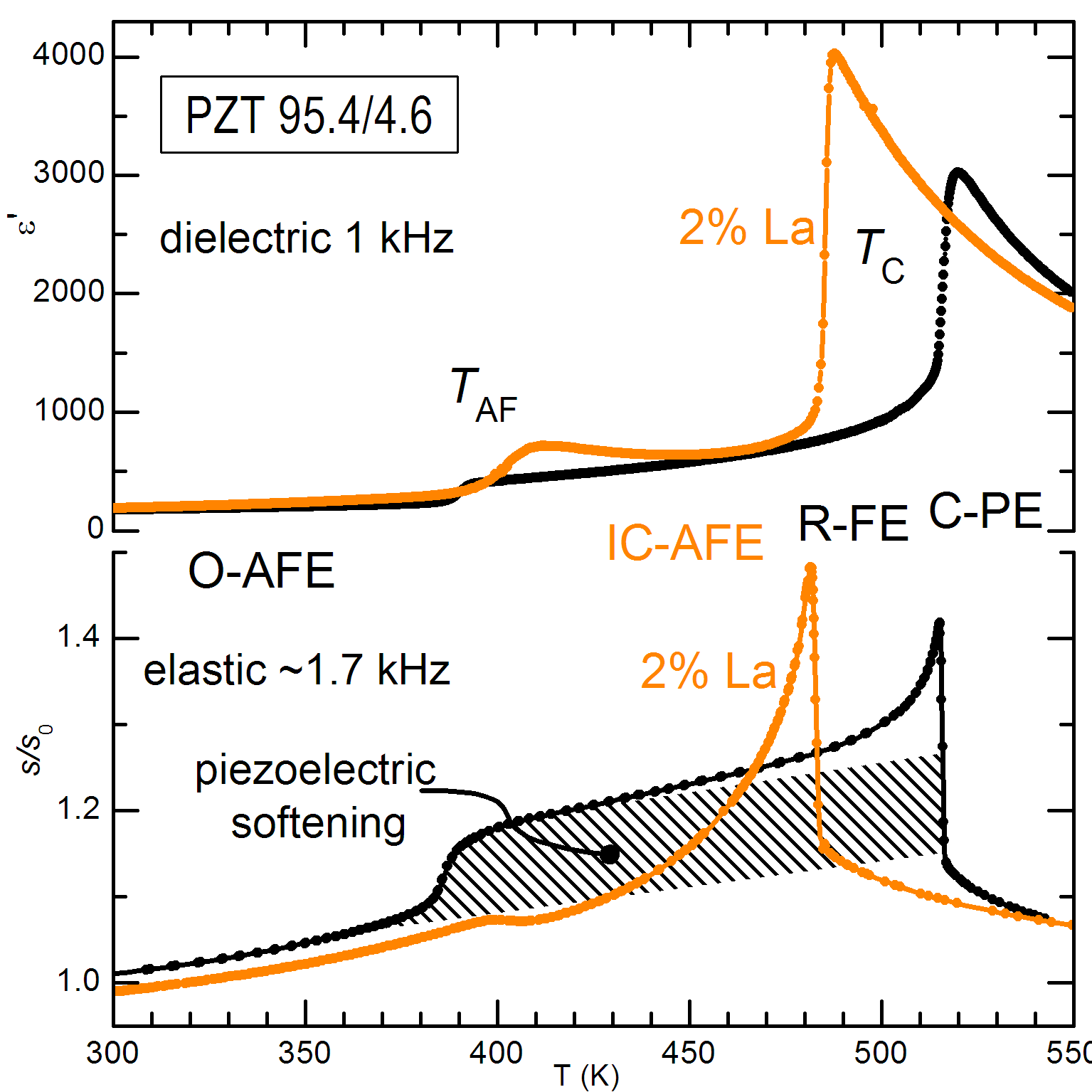}
\caption{Dielectric permittivity $\protect\varepsilon $ and compliance $s$
of PLZT $100x/95.4/4.6$ with $x=0$ and $0.02$ measured during heating. The
ranges of the various phases are indicated. The dashed region is
identifiable with the piezoelectric softening in the FE phase.}
\label{fig_AnDiPiezo}
\end{figure}

The piezoelectric softening is not generally known with this name and is not
always evident as such in the curves of the elastic compliance or modulus
versus temperature. In fact, it may be masked by precursor and fluctuation
effects near the transition, by the domain wall motion, or the occurrence of
other transitions at lower temperature. The effect is instead evident in the
pair of PLZT $100x/95.4/4.6$ compositions in Fig. \ref{fig_AnDiPiezo}. In
fact, apart from the precursor softening approaching $T_{\mathrm{C}}$ from
the PE phase and the peaked contribution just below $T_{\mathrm{C}}$, which
is common to both compositions, a roughly constant softening, shaded in Fig. %
\ref{fig_AnDiPiezo}, is visible only within the FE phase of the La free
sample. The softening abruptly disappears when the R-FE phase transforms
into O-AFE and never appears in the La-doped sample, where also the
intermediate phase is AFE, instead of FE. The $s\left( T\right) $ curve of
undoped sample itself and even more the comparison with $s\left( T\right) $
with La-doping clearly indicates that a roughly temperature independent
softening is associated with the FE state, indicated as a hatched region in
Fig. \ref{fig_AnDiPiezo}.

According to the Landau theory of phase transitions, the standard treatment
of the elastic anomaly at a FE transition\cite{SL98} consists in
establishing an expansion $G_{0}\left( \mathbf{P}\right) $ of the free
energy in powers of the polarization $\mathbf{P}$ with enough terms to
reproduce the features of the relevant region of the phase diagram. This $%
G_{0}$ is inserted in place of $U-TS$ in the elastic Gibbs energy\cite{LG77}
\begin{equation}
G_{1}=U-TS-\mathbf{\sigma }:\mathbf{\varepsilon }  \label{G1-0}
\end{equation}%
so that
\begin{equation}
G_{1}=G_{0}\left( \mathbf{P}\right) -\frac{1}{2}\mathbf{\sigma }:\mathbf{s}%
^{0}:\mathbf{\sigma }-\mathbf{\sigma }:\mathbf{Q:PP}  \label{G1}
\end{equation}%
where $\mathbf{\sigma }$ and $\mathbf{\varepsilon }$ with components $\sigma
_{ij}$ and $\varepsilon _{ij}$ are the stress and strain tensors, $\mathbf{s}%
^{0}$ with components $s_{ijkl}^{0}$ is the bare compliance in the absence
of polarization, hence often indicated as $\mathbf{s}^{D}$ where $\mathbf{D}%
= $ $\mathbf{P}+\mathbf{\epsilon \cdot E}$, $\mathbf{Q}$ with components $%
Q_{ijkl}$ is the electrostrictive coefficient, $\mathbf{PP}$ is a dyad with
components $P_{i}P_{j}$; a single dot indicates scalar product (summation
over repeated indices) over one index, while the semicolon indicates
summation over a pair of indices. For full derivations see the Supplemental
Material. The differential of (\ref{G1-0}) is $dG_{1}=-SdT-\mathbf{%
\varepsilon }:d\mathbf{\sigma }+\mathbf{E}\cdot d\mathbf{D}$, so that the
strain is $\mathbf{\varepsilon }=$ $-\partial G_{1}/\partial \mathbf{\sigma .%
}$ The compliance $\mathbf{s}$ softened by the change of $\mathbf{P}\left(
\mathbf{\sigma }\right) $ under the application of the measuring stress is
often called $\mathbf{s}^{E},$ being the one measured in the absence of an
electric field, that also would change $\mathbf{P}$; it can be calculated as
\begin{eqnarray}
\mathbf{s}^{E} &=&\frac{d\mathbf{\varepsilon }}{d\mathbf{\sigma }}=\frac{d}{d%
\mathbf{\sigma }}\left( -\frac{\partial G_{1}}{\partial \mathbf{\sigma }}%
\right) =\frac{d}{d\mathbf{\sigma }}\left( \mathbf{s}^{0}:\mathbf{\sigma }+%
\mathbf{Q:PP}\right) =  \notag \\
&=&~\mathbf{s}^{0}+2\mathbf{Q:}\left( \mathbf{P}\frac{\mathbf{\partial P}}{%
\mathbf{\partial \sigma }}\right) \mathbf{~}.  \label{sE-1}
\end{eqnarray}%
Usually the derivatives are explicitly calculated for specific expansions $%
G_{0}\left( \mathbf{P}\right) $. In the simplest possible case

\begin{equation}
G_{0}\left( P\right) =\frac{a}{2}P^{2}+\frac{b}{4}P^{4}  \label{G0}
\end{equation}%
with $a=\alpha \left( T-T_{\mathrm{C}}\right) $, the temperature
dependencies of the spontaneous polarizations and their stress derivatives
cancel each other, and one obtains\cite{SL98,Cor15}
\begin{eqnarray}
\mathbf{s}^{E}\left( T>T_{\mathrm{C}}\right) &=&\mathbf{s}^{0} \\
\mathbf{s}^{E}\left( T<T_{\mathrm{C}}\right) &=&\mathbf{s}^{0}+\frac{2%
\mathbf{QQ}}{b}  \label{s-simple}
\end{eqnarray}%
namely a steplike softening below $T_{\mathrm{C}}$. Actually, $Q_{\alpha
}Q_{\beta }$ might be negative for $s_{\alpha \beta }$ with $\alpha \neq
\beta $ ($\alpha \equiv ij$ in Voigt notation,\cite{Nye57} see Supplemental Material,
and with a scalar order parameter $Q_{\alpha ij}\rightarrow Q_{\alpha }$),
resulting in a stiffening, but this eventuality does not seem to be
important in ferroelectric perovskites, where the major effects are in the $%
s_{\alpha \alpha }$ components. For expansions $G_{0}\left( P\right) $
including higher powers of $\mathbf{P}$ and anisotropic terms the expression
of the softening becomes far more complicated than Eq. (\ref{s-simple}), but
for our purposes $G_{0}$ can be left general, and we will rather express $%
\partial \mathbf{P}/\partial \mathbf{\sigma }$ in terms of the piezoelectric
constitutive relations\cite{Dam98,LG77}
\begin{gather}
\mathbf{\varepsilon =s}^{E}\mathbf{:\sigma +d}^{+}\cdot \mathbf{E}
\label{e=ss+dE} \\
\mathbf{D=d}:\mathbf{\sigma }+\mathbf{\epsilon }^{\sigma }\cdot \mathbf{E}
\label{D=ds+eE}
\end{gather}%
where $\mathbf{\epsilon }^{\sigma }$ is the dielectric permittivity at
constant stress and the cross indicates conjugate tensor or transpose
matrix. From Eq. (\ref{D=ds+eE}) with $E=0$ we see that $\mathbf{P}=$ $%
\mathbf{d:\sigma }$ is the piezoelectric charge and $\frac{\partial \mathbf{P%
}}{\partial \mathbf{\sigma }}=\frac{\partial \mathbf{D}}{\partial \mathbf{%
\sigma }}=\mathbf{d}$ is the piezoelectric coefficient. Derivating the
equilibrium condition $0=\frac{\partial G_{1}}{\partial \mathbf{P}}$ with
respect to $\mathbf{\sigma }$ we can also express\cite{Dam98,suppl} $\mathbf{%
Q}$ in terms of $\mathbf{d}\ $%
\begin{equation}
\mathbf{d}=\frac{\partial \mathbf{P}}{\partial \mathbf{\sigma }}=2\mathbf{%
\epsilon \cdot Q\cdot P~,}  \label{dQ}
\end{equation}%
which expresses the well known fact that the piezoelectricity ($d$) is
electrostriction ($Q$) biased by the spontaneous polarization, so that
finally the piezoelectric softening in Eq. (\ref{sE-1}) can be written as%
\begin{equation}
\Delta \mathbf{s}^{\text{soft}}=\mathbf{s}^{E}-\mathbf{s}^{0}=4\left(
\mathbf{\mathbf{Q}\cdot \mathbf{P}}\right) \cdot \mathbf{\epsilon \cdot }%
\left( \mathbf{Q\cdot P}\right) =\mathbf{d}^{+}\cdot \mathbf{\epsilon }%
^{-1}\cdot \mathbf{d}  \label{s-soft}
\end{equation}

Expressions equivalent to Eq. (\ref{s-soft}) have been used to describe the
elastic response of deformation modes that are piezoelectric also in the PE
phase, like the elastic constant $c_{66}$ in KDP,\cite{Mas46,JS62} but this
formulation is general and clearly shows that, independently of the
complications of the FE transformation, the origin of the additional
softening in the FE phase is electrostrictive or equivalently piezoelectric.
In a simple case like Eq. (\ref{G0}) it is $P^{2}\propto $ $\epsilon
^{-1}\propto $ $T$, so that their temperature dependences cancel out and one
remains with a constant softening in the FE phase, as in (\ref{s-simple}),
but in general some temperature dependence can be expected, besides the
fluctuation effects. Therefore, the piezoelectric contribution hatched in
Fig. \ref{fig_AnDiPiezo} is constant, but might well exhibit some
enhancement on approaching $T_{\mathrm{C}}$.

The $\mathbf{s}^{E}$ compliance is measured when neither an external nor an
internal depolarizing field $\mathbf{E}^{\text{dep}}$ affects or neutralizes
the piezoelectric strain. This is certainly the situation of resonant and
subresonant measurements of unpoled ceramic samples, where the direction of
the spontaneous $\mathbf{P}_{0}$ and the associated depolarization field
changes randomly many times within the regions where $\mathbf{P}$
accumulates the piezoelectric charges (electrodes and vibration nodes) and
averages to zero. In the same situation is a poled crystal or ceramic whose
electrodes are shorted or driven by a circuit with sufficiently low
impedance. In these cases the complete piezoelectric softening Eq. (\ref%
{s-soft}) is observed\ (see \textit{e.g.} the coincidence of the
piezoelectric softening measured in BCTZ with Dynamic Mechanical Analyzer,
electrostatically excited free flexural resonance and piezoresonance\cite%
{CCD14b}), but otherwise the well known phenomenon of piezoelectric
stiffening\cite{Mas39,Mas46,HW62,Aul73,Mac75,DR80b,Ike82,EC87} occurs, which
we show now can exactly cancel the piezoelectric softening, leaving the bare
elastic constant.

In the absence of mechanisms for neutralizing the piezoelectric charges $%
\delta \mathbf{P=d}:\mathbf{\sigma }$ (\textit{e.g.} vibration of a poled
sample with open electrodes), we are in the condition of constant $\mathbf{D}%
=\mathbf{\epsilon }\cdot \mathbf{E}+\mathbf{P}$, where a change $\delta
\mathbf{P}$ causes a depolarization field $\delta \mathbf{E}^{\text{dep}%
}=-\delta \mathbf{P}/\epsilon $ that, according to the additional
constitutive relationships\cite{Dam98,LG77}
\begin{gather}
\mathbf{\varepsilon =s}^{D}\mathbf{:\sigma +g}^{+}\cdot \mathbf{D} \\
\mathbf{E=}-\mathbf{g}:\mathbf{\sigma }+\mathbf{\epsilon }^{-1}\cdot \mathbf{%
D}
\end{gather}%
can be written as $\mathbf{E}^{\text{dep}}=-\mathbf{g:\sigma }$. and, since%
\cite{suppl} $\mathbf{g}=\mathbf{\epsilon }^{-1}\cdot \mathbf{d}$,
\begin{equation}
\mathbf{E}^{\text{dep}}=-\mathbf{\epsilon }^{-1}\cdot \mathbf{\mathbf{d}%
:\sigma ~.}
\end{equation}%
This field causes an opposing stress with respect to the constant $E$
condition, and its effect can be written with Eq. (\ref{e=ss+dE}) as
\begin{equation}
\mathbf{\varepsilon }=\mathbf{s}^{E}:\mathbf{\sigma }-\mathbf{d}^{+}\cdot
\mathbf{\ \mathbf{\epsilon }^{-1} \cdot \mathbf{d}:\sigma =s}^{D}\mathbf{%
:\sigma }
\end{equation}%
in terms of a reduced and hence restiffened compliance $\mathbf{s}^{D}$.
Therefore the relationship between compliance at constant $E$ and $D$ is
\begin{equation}
\Delta \mathbf{s}^{\text{stiff}}=\mathbf{s}^{D}-\mathbf{s}^{E}=-\mathbf{d}%
^{+}\cdot \mathbf{\mathbf{\epsilon }^{-1}\cdot d=-}\Delta \mathbf{s}^{\text{%
soft}}~;  \label{s-stiff}
\end{equation}
the piezoelectric stiffening simply cancels the piezoelectric softening and
one remains with the compliance of the PE phase $\mathbf{s}^{0}\equiv
\mathbf{s}^{D}$.

Piezoelectric stiffening is known to occur in resonating devices\cite{Ike82}
and to affect the propagation of elastic waves.\cite%
{HW62,Aul73,Mac75,DR80b,EC87} Moreover, if the material is conducting, the
free charge may partially neutralize the depolarization fields and yield a
dispersion in frequency of the relaxation type, with intermediate elastic
constant between the completely softened or restiffened values.\cite{HW62}

In the case of the propagation of acoustic waves, the piezoelectric
stiffening fully acts in a polarized material or even in unpoled ceramics if
the wavelength is smaller than the size of the FE domains, as for Brillouin
scattering. In both cases the depolarization field fully develops within
each half wavelength within domains of uniform polarization. In this context
the piezoelectrically softened elastic constant $\mathbf{c}^{E}=\left(
\mathbf{s}^{E}\right) ^{-1}$ is considered the reference, and expressions
are provided for the restiffened constants $\mathbf{c}^{D}=$ $\mathbf{c}%
^{E}+\Delta \mathbf{c}^{\text{stiff}}$ felt by a plane wave propagating
along the direction $\mathbf{\hat{n}}$\cite{Aul73,Mac75,DR80b,EC87}
\begin{equation}
\Delta \mathbf{c}^{\text{stiff}}\left( \mathbf{\hat{n}}\right) =\frac{\left(
\mathbf{e\cdot \hat{n}}\right) \left( \mathbf{e\cdot \hat{n}}\right) }{%
\mathbf{\hat{n}\cdot \epsilon \cdot \hat{n}}}  \label{c-stiff}
\end{equation}%
where the relationship between piezoelectric stress and strain coefficients
is\cite{suppl} $\mathbf{e}=\mathbf{d:c}^{E}$. This formula can be rewritten
in a manner directly comparable to Eq. (\ref{s-stiff}); for special
directions where only one of the tensor components projected onto $\mathbf{%
\hat{n}}$ is relevant,
\begin{equation}
\frac{\Delta c^{\text{stiff}}}{c^{E}}=\frac{\left( d~c^{E}\right) ^{2}}{%
c^{E}~\epsilon }=\frac{d^{2}}{s^{E}~\epsilon }=\frac{\mathbf{-}\Delta
\mathbf{s}^{\text{soft}}}{s^{E}}~.  \label{cs-ss}
\end{equation}%
Also for propagating acoustic waves, the piezoelectric softening may be
canceled by the formation of the depolarization field.

\begin{figure}[tbh]
\includegraphics[width=8.5 cm]{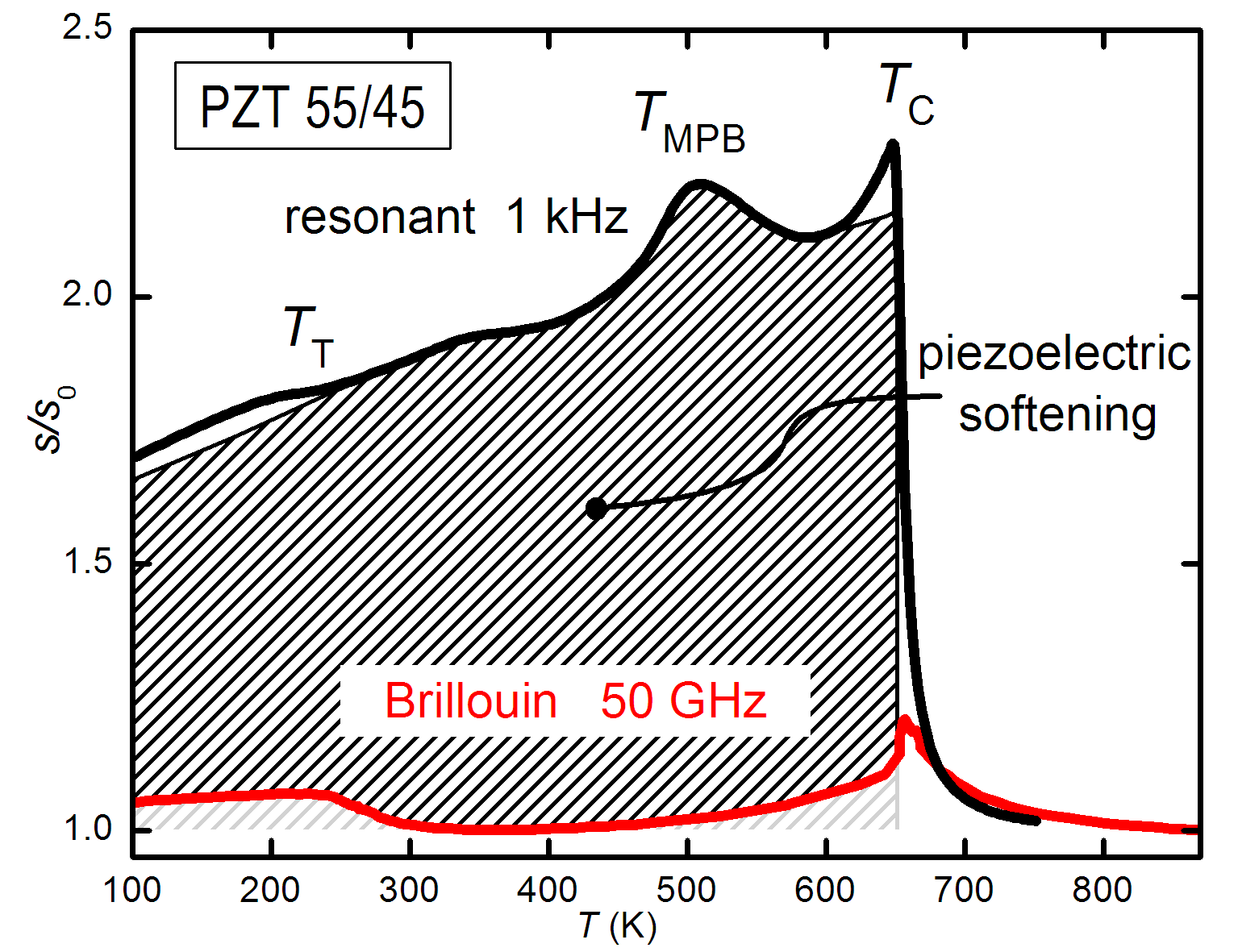}
\caption{Normalized compliance of PZT $55/45$ measured with flexural
resonance\protect\cite{CTC13} at 1~kHz and Brillouin scattering\protect\cite%
{KKK12} at 50~GHz.}
\label{fig_res-BS}
\end{figure}

Figure \ref{fig_res-BS} presents an example where the clear piezoelectric
softening measured at low frequency on a ceramic PZT sample is totally
absent in the Brillouin scattering experiment. At the PZT 55/45 composition
one crosses at $T_{\mathrm{MPB}}$ the morphotropic phase boundary (MPB)
between R and tetragonal (T) FE phases, and further below $T_{\mathrm{T}}$
the oxygen octahedra undergo long range tilting about the polar axis, which
is a non polar mode. Accordingly, the compliance measured at $\sim 1$~kHz%
\cite{CTC13} with the same method as described here has a large jump below $%
T_{\mathrm{C}}$, where the hatched region is considered piezoelectric
softening, and is followed by a peak at $T_{\mathrm{MPB}}$. As discussed
elsewhere,\cite{127,Cor15,CLM16} this peak is due to an enhancement of the
shear compliance $\Delta s_{55}$, that is approximately linearly coupled to
the rotation of the polarization between the T $\left[ 001\right] $ and the
R $\left[ 111\right] $ axis. Since this $\Delta s_{55}$ is directly related
to the FE state, it contributes to the piezoelectric coupling\cite{CLM16}
and is totally considered piezoelectric softening. Instead, the step below $%
T_{\mathrm{T}}$ is not considered piezoelectric softening, because it is due
to the antiferrodistortive octahedral tilting, which does not directly
contribute to the polarization state. The lower curve in Fig. \ref%
{fig_res-BS} is extracted as $s/s_{0}=$ $\left( \nu _{0}/\nu _{\mathrm{B}%
}\right) ^{2}$ from the Brillouin frequency shift $\nu _{\mathrm{B}}\sim 50$%
~GHz for the LA mode along $\left[ 100\right] $, as measured by Kojima
\textit{et al.}\cite{KKK12} In this geometry the $c_{11}$ constant is
probed, which is responsible for the jump below $T_{\mathrm{C}}$ in the
previous measurement, but the probed acoustic waves develop a longitudinal
piezoelectric polarization that is completely canceled by the depolarization
field and consequent piezoelectric stiffening, as in Eq. (\ref{cs-ss}).
Therefore, the piezoelectric softening is totally absent, including the
enhancement at $T_{\mathrm{MPB}}$ related to $s_{55}$. On the other hand,
the peaked softening due to the short range polar fluctuations near $T_{%
\mathrm{C}}$ is clearly visible, because their correlation lengths are
shorter than the Brillouin wavelength and there is no coherent
depolarization field within a half wavelength. Also the the step below $T_{%
\mathrm{T}}$ is fully visible, because it does not involve piezoelectricity.

The smallness of the softening in the Brillouin experiments compared to
those at lower frequency has been noticed also for other FE materials, and
explained in terms of dynamic response of the order parameter and domain
wall motion\cite{Wie09}, but besides relaxation and adiabatic versus
isothermal effects, the piezoelectric stiffening from the depolarization
field should also be taken into account.

\section{Conclusion}

It is shown that, regardless of the possibly complicated phase diagram,
within a FE phase the elastic constants undergo a softening due to the
additional piezoelectric strain under the application of the probing stress.
Such a strain can be reduced or canceled by the development of the
depolarization field $\mathbf{E}^{\text{dep}}$ due to the piezoelectric
charges, through the well known mechanism of the piezoelectric stiffening,
which therefore restores the bare elastic constants of the paraelectric
phase. The full piezoelectric softening can be observed in resonant and
subresonant measurements of unpoled ceramic samples, where no coherent
macroscopic $\mathbf{E}^{\text{dep}}$ can develop, or also of poled samples
where $\mathbf{E}^{\text{dep}}$ is shorted by an external circuit. Instead,
in Brillouin scattering experiments, where the wavelength of the probed
acoustic waves is smaller than the domain size, a coherent $\mathbf{E}^{%
\text{dep}}$ forms within each half wavelength and cancels the piezoelectric
softening. It follows that the compliance curves versus temperature of
unpoled ceramic samples indirectly probe the piezoelectric coupling, and for
example allow to distinguish FE from AFE phases and to identify temperature
and composition regions of particularly high piezoelectricity.

Two experimental verifications are presented. Elasticity and dielectric
measurements are carried out in PZT compositions where the FE phase turns
into AFE by cooling or doping with La, with concomitant restiffening. In
addition, for PZT 55/45 a comparison is discussed of the fully
piezoelectrically softened Young's modulus with the unsoftened $c_{11}$
elastic constant from the Brillouin scattering experiments of Kojima and
coworkers.\cite{KKK12}

\begin{acknowledgments}
The authors thank P.M. Latino (CNR-ISC) for his technical assistance in the
realization of the anelastic experiments and C. Capiani (CNR-ISTEC) for the
skillful preparation of the samples.
\end{acknowledgments}


\numberwithin{equation}{section}
\section{Supplemental Material}

The various equations used in the text are written out in terms of their
components. Indices in Latin letters are the cartesian components $i=1-3$ of
the electrical quantities, while the Greek letters are the stress/strain
components in Voigt or matrix notation:\cite{Nye57} $ij=11,22,33\rightarrow
\alpha =1,2,3$, $ij=12,21\rightarrow \alpha =6$, $ij=23,32\rightarrow \alpha
=4$, $ij=13,31\rightarrow \alpha =5$, with the additional rule that the
components of $\mathbf{\varepsilon }$, $\mathbf{s}$ and $\mathbf{Q}$ have to
be multiplied by 2 for each index $\alpha \geq 4$. The convention is adopted
of summation over repeated indices, \textit{e.g.} $\sigma _{\alpha
}Q_{\alpha ij}P_{i}P_{j}\equiv \sum_{\alpha
=1}^{6}\sum_{i=1}^{3}\sum_{j=1}^{3}\sigma _{\alpha }Q_{\alpha ij}P_{i}P_{j}$

\subsection{Piezoelectric constitutive relations}

The piezoelectric constitutive relations (Eqs. (7,8)\ and (11,12) in the
article) are:\cite{LG77,Dam98}
\begin{gather}
\varepsilon _{\alpha }=s_{\alpha \beta }^{E}\sigma _{\beta }+d_{j\alpha
}E_{j} \\
D_{i}=d_{i\beta }\sigma _{\beta }+\epsilon _{ij}^{\sigma }E_{j}
\label{D=ds+eE}
\end{gather}%
where $d_{j\alpha }=$ $d_{\alpha j}^{+}$ and
\begin{gather}
\varepsilon _{\alpha }=s_{\alpha \beta }^{D}\sigma _{\beta }+g_{j\alpha
}D_{j} \\
E_{i}=-g_{i\beta }\sigma _{\beta }+\beta _{ij}^{\sigma }D_{j} \\
\text{with }\beta _{ik}\epsilon _{kj}=\delta _{ij}
\end{gather}%
The relationship between the piezoelectric constants $\mathbf{d}$ and $%
\mathbf{g}$ can be found as
\begin{eqnarray}
g_{i\alpha } &=&\left( \frac{\partial \varepsilon _{\alpha }}{\partial D_{i}}%
\right) _{\sigma }=\left( \frac{\partial \varepsilon _{\alpha }}{\partial
E_{j}}\frac{\partial E_{j}}{\partial D_{i}}\right) _{\sigma }= \\
&=&d_{j\alpha }\beta _{ij}^{\sigma }=\left( \epsilon ^{-1}\right)
_{ij}^{\sigma }d_{j\alpha }
\end{eqnarray}

When dealing with the piezoelectric stiffening of the elastic constant $%
\mathbf{c=}$ $\mathbf{s}^{-1}$, one makes use of another pair of
constitutive relations\cite{Dam98}
\begin{eqnarray}
\sigma _{\alpha } &=&c_{\alpha \beta }^{E}\varepsilon _{\beta }-e_{j\alpha
}E_{j}  \label{pzss1} \\
D_{i} &=&e_{i\beta }\varepsilon _{\beta }+\epsilon _{ij}E_{j}
\label{pzss2}
\end{eqnarray}%
and the relationship between the piezoelectric coefficients $\mathbf{e}$ and
$\mathbf{d}$ is
\begin{equation}
e_{i\alpha }=\left( \frac{\partial D_{i}}{\partial \varepsilon _{\alpha }}%
\right) _{E}=\left( \frac{\partial D_{i}}{\partial \sigma _{\beta }}\right)
_{E}\left( \frac{\partial \sigma _{\beta }}{\partial \varepsilon _{\alpha }}%
\right) _{E}=d_{i\beta }c_{\alpha \beta }^{E}
\end{equation}

\subsection{Compliance in the absence of depolarization field}

The differential of the internal energy is\cite{LG77}

\begin{equation}
dU=TdS+\sigma _{j}d\varepsilon _{j}+E_{j}dD_{j}
\end{equation}%
with $S=$ entropy, $\mathbf{\varepsilon }$ and $\mathbf{D}$ as independent
variables. Since we need as independent variables $T$ and $\mathbf{\sigma },$
we use the elastic Gibbs energy
\begin{equation}
G_{1}=U-TS-\sigma _{\alpha }\varepsilon _{\alpha }
\end{equation}%
whose differential is%
\begin{equation}
dG_{1}=-SdT-\varepsilon _{\alpha }d\sigma _{\alpha }+E_{j}dD_{j}
\label{dG1}
\end{equation}

In the absence of external fields one can deduce the phase transitions and
phase diagram of the material by a suitable $G=U-TS$, which, applying the
Landau theory of phase transitions to a ferroelectric, is expanded in powers
of $\mathbf{P}$ as $G_{0}\left( P\right) $. Of the additional terms of the
expansion of the free energy in powers of the variables $P$, $T$ and $\sigma
$ we explicitly take into account only the elastic ($\propto \sigma _{\alpha
}\sigma _{\beta }$) and electrostrictive ($\propto \sigma _{\alpha
}P_{i}P_{j}$), since the piezoelectric ($\propto \sigma _{\alpha }P_{i}$) is
generally absent in the paraelectric phases:

\begin{equation}
G_{1}=G_{0}\left( P\right) -\frac{1}{2}\sigma _{\alpha }s_{\alpha
\beta }^{0}\sigma _{\beta }-\sigma _{\alpha }Q_{\alpha ij}P_{i}P_{j}
\label{G1}
\end{equation}%
where $\mathbf{s}^{0}$ is the bare elastic compliance in the absence of
electrostrictive and hence piezoelectric effect. The compliance that is
measured when $\mathbf{P}$ can freely reach equilibrium with stress in the
absence of external field is
\begin{equation}
s_{\alpha \beta }^{E}=\frac{d\varepsilon _{\alpha }}{d\sigma _{\beta }}~,
\end{equation}%
where the strain can be deduced from Eqs. (\ref{dG1},\ref{G1})
\begin{equation}
\varepsilon _{\alpha }=-\frac{\partial G_{1}}{\partial \sigma _{\alpha }}%
=s_{\alpha \beta }^{0}\sigma _{\beta }+Q_{\alpha ij}P_{i}P_{j}\mathbf{~},
\end{equation}%
where $\mathbf{Q:PP}$ is the electrostrictive strain, that in the FE phase
with $\mathbf{P=P}_{0}\neq 0$ becomes a piezoelectric strain. The compliance
at constant $\mathbf{E}$, with $\mathbf{P}$ and $\mathbf{D}$ in equilibrium
with the applied stress is:%
\begin{equation}
s_{\alpha \beta }^{E}=\frac{d\varepsilon _{\alpha }}{d\sigma _{\beta }}%
=s_{\alpha \beta }^{0}+2Q_{\alpha ij}P_{i}\frac{\partial
P_{j}}{\partial \sigma _{\beta }} \label{sE}
\end{equation}

If $E=0$, we can deduce from Eq. (\ref{D=ds+eE})
\begin{equation}
\frac{\partial P_{i}}{\partial \sigma _{\alpha }}=\frac{\partial D_{i}}{%
\partial \sigma _{\alpha }}=d_{i\alpha }  \label{d}
\end{equation}%
and can express $\mathbf{Q}$ in terms of $\mathbf{d}$. In fact, let us write
the condition for the equilibrium (spontaneous) polarization

\begin{equation}
0=\frac{\partial G_{0}}{\partial P_{i}}-2\sigma _{\beta }Q_{\beta ij}P_{j}
\end{equation}%
and further derivate with respect to $\sigma _{\alpha }$, taking into
account that $G_{0}$ depends on $\mathbf{\sigma }$ through $\mathbf{P}$ and
taking the limit $\sigma \rightarrow 0$%
\begin{eqnarray*}
0 &=&\frac{d}{d\sigma _{\alpha }}\left( \frac{\partial G_{0}}{\partial P_{i}}%
-2\sigma _{\beta }Q_{\beta ij}P_{j}\right) = \\
&=&\frac{\partial ^{2}G_{0}}{\partial P_{i}\partial P_{j}}\frac{\partial
P_{j}}{\partial \sigma _{\alpha }}-2Q_{\alpha ij}P_{j}-O\left( \sigma
\right)
\end{eqnarray*}%
where the dielectric stiffness is
\begin{equation}
\frac{\partial ^{2}G_{0}}{\partial P_{i}\partial P_{j}}=\beta _{ij}=\left(
\epsilon ^{-1}\right) _{ij}
\end{equation}%
and multiplying by $\mathbf{\epsilon }$ and using (\ref{d})%
\begin{equation}
\frac{\partial P_{i}}{\partial \sigma _{\alpha }}=2\epsilon
_{ik}Q_{\alpha kl}P_{l}=d_{i\alpha }  \label{d-dPds}
\end{equation}%
so that finally the piezoelectric softening can be written as
\begin{eqnarray}
\Delta s_{\alpha \beta }^{\text{soft}} &=&s_{\alpha \beta }^{E}-s_{\alpha
\beta }^{0}=2Q_{\alpha ij}P_{i}\frac{\partial P_{j}}{\partial \sigma _{\beta
}}=  \notag \\
&=&2Q_{\alpha ij}P_{i}2\epsilon _{jk}Q_{\beta kl}P_{l}
\end{eqnarray}%
and exploiting the symmetry $Q_{\alpha ij}=Q_{\alpha ji}$ we obtain the
first part of Eq. (10)\ in the article
\begin{equation}
\Delta s_{\alpha \beta }^{\text{soft}}=4Q_{\alpha ji}P_{i}\epsilon
_{jk}Q_{\beta kl}P_{l}
\end{equation}%
that can be further expressed in terms of the piezoelectric coupling $%
\mathbf{d}$\textbf{\ }using (\ref{d-dPds}) and inserting $\mathbf{\epsilon }=
$ $\mathbf{\epsilon \cdot }$ $\mathbf{\epsilon }^{-1}\cdot \mathbf{\epsilon }
$ or $\epsilon _{jk}=$ $\epsilon _{jm}\left( \epsilon ^{-1}\right)
_{mn}\epsilon _{nk}$ with $\epsilon _{jm}=\epsilon _{mj}$ and $d_{m\alpha }=$
$d_{\alpha m}^{+}$
\begin{eqnarray}
\Delta s_{\alpha \beta }^{\text{soft}} &=&2\epsilon _{mj}Q_{\alpha
ji}P_{i}\left( \epsilon ^{-1}\right) _{mn}2\epsilon _{nk}Q_{\beta kl}P_{l}=
\notag \\
&=&d_{\alpha m}^{+}\left( \epsilon ^{-1}\right) _{mn}d_{n\beta }
\end{eqnarray}%
which is the last part of Eq. (10)\ in the article.

\end{document}